# Beyond the Binary: The System of All-round Evaluation of Research and Its Practices in China

Yu Zhu*, Jiyuan Ye**

* *zhu.yu@smail.nju.edu.cn*
0000-0002-2548-828X
School of Information Management, Nanjing University, China

** *yejiyuan@nju.edu.cn*
0000-0002-4232-8923
School of Information Management, Nanjing University, China

The lack of a macro-level, systematic evaluation theory to guide the implementation of evaluation practices has become a key bottleneck in the reform of global research evaluation systems. By reviewing the historical development of research evaluation, this paper highlights the current binary opposition between qualitative and quantitative methods in evaluation practices. This paper introduces the System of All-round Evaluation of Research (SAER), a framework that integrates form, content, and utility evaluations with six key elements. SAER offers a theoretical breakthrough by transcending the binary, providing a comprehensive foundation for global evaluation reforms. The comprehensive system proposes a trinity of three evaluation dimensions, combined with six evaluation elements, which would help academic evaluators and researchers reconcile binary oppositions in evaluation methods. The system highlights the dialectical wisdom and experience embedded in Chinese research evaluation theory, offering valuable insights and references for the reform and advancement of global research evaluation systems.

## 1. Introduction

With the rapid growth of China's scientific research output and its increasing share in global research (Tollefson, 2018), the reform of the research evaluation system has become a focal point for both Chinese research management departments and the academic community. As early as 2010, scholars had already identified persistent issues within the academic evaluation system, including excessive quantification, overemphasis on formalities, excessive administrative intervention, weakened evaluation of innovation, favoritism among review experts, and compromised evaluation results (Ye, 2010b). These issues not only contribute to academic misconduct and corruption but also, to some extent, hinder the long-term sustainable and healthy development of academic research.

Since 2020, China have explicitly advocated for moving away from "SCI Supremacy" and shifting the evaluation system from focusing on quantitative and formal indicators to emphasizing academic content and practical contributions, aiming to return academic evaluation to its essence of quality assessment (Ye, 2020b). This reform initiative has also sparked widespread global discussion (Editorial in Nature, 2020; Mallapaty, 2020), highlighting the international academic community's attention to China's reforms in its research evaluation system. Meanwhile, despite the launch of international declarations such as *the San Francisco Declaration on Research Assessment* (DORA, 2012) and *the Leiden Manifesto* (Hicks et al., 2015), which advocate for reform in academic evaluation, the deep-rooted theoretical issues within the current academic evaluation system remain unresolved. Therefore, constructing a rational evaluation system to clarify ambiguities in the field, define the direction and pathways for evaluation practice, and provide a robust methodological foundation for research evaluation remains a pressing issue for both the global academic community and research management institutions.

In fact, in 2010, the Chinese academic evaluation community has proposed a systematic and forward-looking academic evaluation theory — the System of All-round Evaluation of

Research (SAER) (Ye, 2010b, 2021) — to address issues such as weak evaluation theory, unclear evaluation subjects, ambiguous evaluation purposes, vague evaluation standards and indicators, singular evaluation methods, incomplete evaluation systems, and low credibility of evaluation results. As a representative of China's academic evaluation theory (Information Center for Social Science Renmin University of China, 2025), SAER strategically offers a holistic solution to address evaluation-related issues, designs a phased and systematic approach for implementation, and fosters a constructive interplay between evaluation theory and practice. This theory framework plays a crucial role in tackling issues such as the overemphasis on quantitative metrics, while also fostering academic integrity, creating an inclusive and supportive research environment, and promoting scientific advancement. Given the above context, this paper focuses on the epistemological aspects of research evaluation theory and introduces relevant practical applications.

## 2. The Binary of Evaluation System

The formation of academic evaluation systems is largely dependent on evaluation methods. Peer review and bibliometric analysis are the two most widely adopted approaches, yet both suffer from inherent limitations, contributing to the current binary landscape of research assessment.

Peer review, as the earliest academic evaluation mechanism, involves independent assessments of scholarly work by experts within the scientific community based on established criteria (Zhang & Sivertsen, 2020). Its institutionalization dates back to the 17th century, marked by the establishment of the Royal Society of London in 1662 and the Académie Royale des Sciences in Paris in 1699, along with their respective journals, *Philosophical Transactions* and *Journal des Sçavans*, which signified a transition from private scholarly correspondence to standardized scientific publishing (Lee et al., 2013). With the evolution of scholarly communication, peer review has diversified into newer forms such as open review and preprints. However, its fundamental reliance on subjective expert judgment renders it vulnerable to biases stemming from individual perspectives and social influences (Marsh et al., 2008) and the absence of robust feedback mechanisms further undermines the reliability of the process (Derrick, 2018). Additionally, the substantial financial and time costs associated with peer review make high-quality evaluations a scarce resource (Technopolis Group, n.d.), struggling to meet the demands of large-scale and rapid scholarly assessment (Abramo & D'Angelo, 2011). Despite these challenges, best practices in peer review have been exemplified by initiatives such as the UK's *Research Excellence Framework* (REF; Higher Education Funding Council for England, n.d.).

On the other hand, bibliometric analysis, a quantitative approach based on publication and citation frequencies, has gained widespread application in research evaluation. This method dates back to 1955 when Eugene Garfield proposed using citation frequencies and impact factors to assess academic journals (Garfield, 1955). While citation data can partially reflect the outcomes of peer review (Ye, 2020b), the indiscriminate treatment of citation motivations in bibliometric analysis (Brooks, 1986) may lead to criticisms regarding citation inflation and manipulation (MacRoberts & MacRoberts, 1989). In response to these concerns, the *San Francisco Declaration on Research Assessment* (DORA, 2012) and the *Leiden Manifesto* (Hicks et al., 2015) both emphasize that research evaluation should be centred on the scientific content of papers rather than journal-based supplementary bibliometric indicators.

Given the strengths and limitations of both mainstream evaluation methods, recent practices have increasingly sought to integrate bibliometric indicators with peer review for a more balanced approach. Examples include the *Strategy Evaluation Protocol* in the Netherlands (Association of Universities in the Netherlands, 2020) and Italy's *Research Evaluation Exercise* (Franceschet & Costantini, 2011), both of which embody the principle of methodological complementarity. This shift signals a growing recognition of the need to move beyond form-driven assessment toward content- and utility-oriented evaluation.

Nevertheless, research on academic evaluation still lacks a systematic theoretical framework, and the persistent dichotomy between qualitative and quantitative approaches remains a dominant paradigm. The SAER offers a theoretical breakthrough by introducing a triadic model that integrates formal evaluation, content evaluation, and utility evaluation. This framework not only provides a holistic perspective for global research evaluation practitioners and scholars but also helps to transcend the binary opposition that has long characterized academic assessment. By offering a more comprehensive theoretical foundation, SAER contributes to a deeper understanding of the historical development, current challenges, and future directions of research evaluation.

### 3. Theoretical Explanation of SAER

The following subsections present assumptions of premises, theoretical foundations, analytical framework, and originality and innovation of SAER.

*3.1 Premise Assumptions and Theories*

**(1) Assumption of premises**
   (a) **Universality assumption**: In all aspects of human activity, individuals continuously make value judgments about various matters related to themselves in order to make choices and take action.
   (b) **Complexity assumption**: Like any other subject, academic research—including its outputs, scholars, and institutions—can be evaluated. However, because academic research concerns the exploration of laws governing phenomena, as well as questions of truth, goodness, and beauty, its evaluation is inherently more complex than that of other domains.
   (c) **Guidance assumption**: Compared to individual assessments, any organized, fair, and reasonable academic evaluation provides more explicit guidance to the evaluated subject, including diagnostic, predictive, selective, and motivational functions.
   (d) **Sufficiency assumption**: The more comprehensive and profound the understanding of the evaluation object, the more reasonable and reliable the evaluation.

**(2) Theoretical Foundations**
   (a) **Value Theory, Epistemology, and Praxis—With a Focus on Marxist Value Theory**

Marxist value theory views value as a social-historical relationship, where an object's value is determined by its ability to meet human needs (Marx & Engels, 1975). It stresses practice as the test for truth and integrate human intent and dialectical objectivity. Therefore, evaluation can reveal objective value and construct a value system, where value judgments go beyond empirical facts to consider the nature of the object, its interaction with the subject, and human needs and ideals. So, unlike truth, value judgments balance plurality and singularity, with "ought" shaped by both objective facts and human aspirations. These perspective lays the foundation for the objectivity standard in SAER.

   (b) **Systems theory, cybernetics, and information theory—with an emphasis on complete information theory**

From a systems theory perspective, objective phenomena can be understood as holistic systems composed of interrelated and interacting elements (Bertalanffy, 1969). Open systems adjust their internal components to achieve new equilibria in response to external environmental changes. As a complex evaluation system, SAER draws directly from systems theory, employing a hierarchical and structured analytical framework to optimize evaluation processes under the principle of system efficiency. Cybernetics conceptualizes system control as an information process (Wiener, 1965), focusing on feedback mechanisms to regulate and control uncertainty, ensuring alignment with predetermined objectives (Shannon, 1948).

SAER's information-theoretic approach is built upon the complete information theory proposed by Zhong (1991). This theory categorizes information into three levels: syntactic information, semantic information, and pragmatic information, corresponding to the form, content, and utility of information, respectively. Under this framework, academic evaluation subjects—such as scholarly outputs, academic institutions, or individual scholars—can be conceptualized as "information entities". This allows for the construction of an evaluation index system or a general evaluation framework based on three dimensions: form, content, and utility.

*3.2 Analytical Framework*
SAER (Ye, 2021) is a conceptual system, a set of application principles, and a framework of rules for academic evaluation as illustrated in Figure 1. The term "All-round" in SAER carries twofold meaning: First, it encompasses all six key elements involved in academic evaluation; second, it considers the entirety of the evaluation object across three dimensions. Specifically, the six key elements of SAER include the evaluation subject, evaluation object, evaluation purpose, evaluation methods, evaluation standards and indicators, and evaluation mechanism. These elements interact to form an integrated evaluation framework. Meanwhile, the three dimensions of evaluation refer to the form, content, and utility of the evaluation object. In essence, SAER is an evaluation system structured around six key elements, three composite evaluation dimensions, and their respective definitions, inferences, explanations, and application principles.

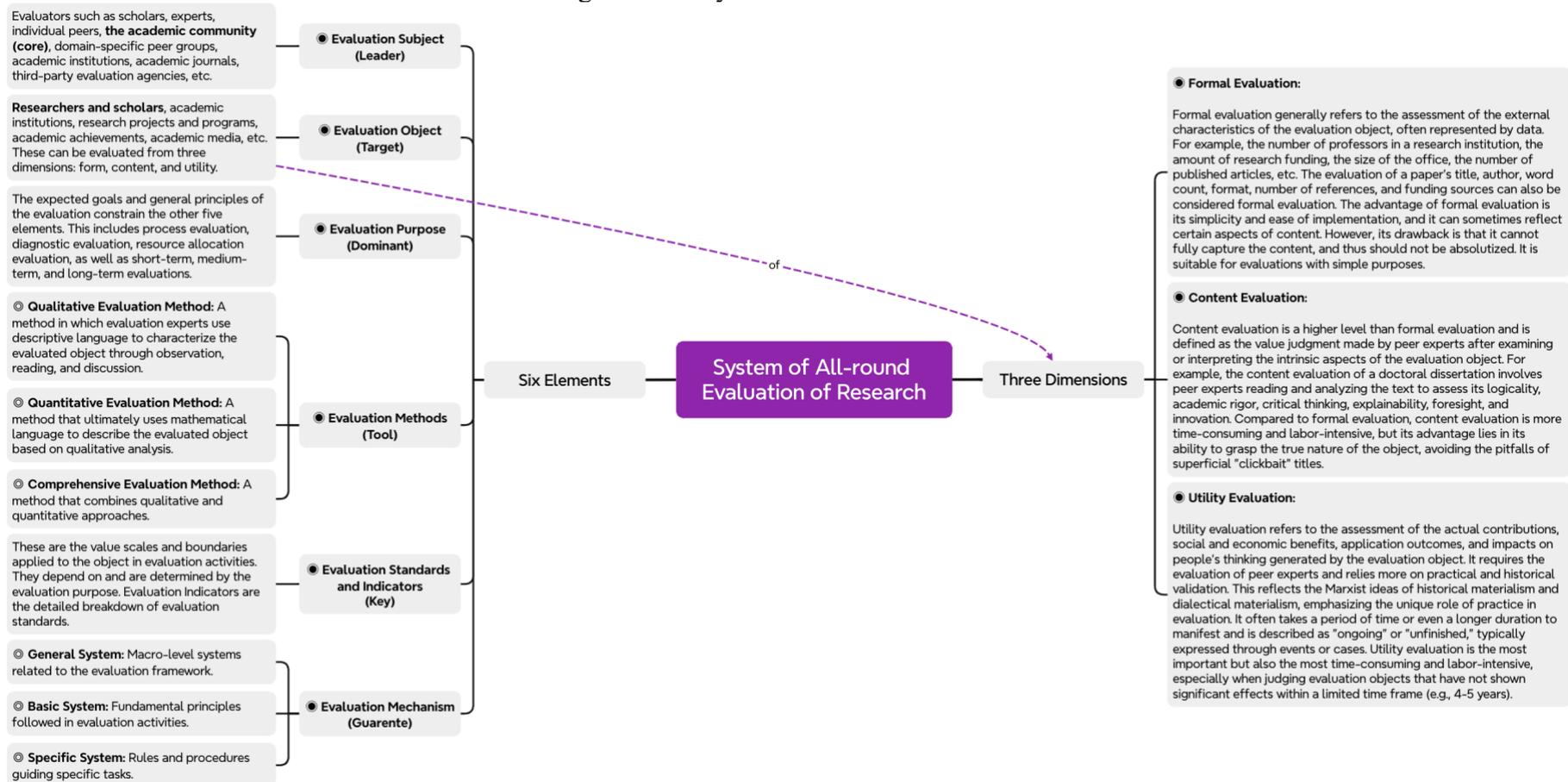

Figure 1: Analytical Framework of SAER.

Among the six elements in Figure 1, the evaluation purpose serves as the guiding principle, shaping the selection of the evaluation subject, the classification of evaluation objects, the choice of evaluation methods, the determination of evaluation standards and indicators, and the design of the evaluation mechanism. In SAER, peer experts play a leading role as the evaluation subject, the evaluation object serves as the foundation, evaluation standards and indicators form the core, evaluation methods provide the means, and the evaluation mechanism ensures institutional support.

From this dialectically unified framework, six main theoretical inferences emerge:
        (1) Evaluation Purpose Constraint Theory,
        (2) Peer Expert Dominant and Meta-Evaluation Theory,
        (3) Evaluation Object Classification Theory,
        (4) Bibliometric Auxiliary Verification Theory,
        (5) Evaluation Standards and Indicators Core and Appropriateness Theory, and
        (6) Evaluation Mechanism Guarantee Theory.

*3.3 Originality and Innovation*
SAER demonstrates originality and innovation in the following areas:
   **(1) Conceptual Innovations**
      (a) Introducing the concept of "All-round Evaluation". The term "All-round" extends beyond the six evaluation elements and three evaluation dimensions in Figure 1, embracing a comprehensive, holistic approach that encompasses the entire process and full life cycle.
      (b) Defining the conceptual integration of formal evaluation, content evaluation, and utility evaluation, incorporating Marxist epistemology, value theory, and praxeology into evaluation theory.
      (c) Differentiating between broad and narrow definitions of academic evaluation, as well as between institutional evaluation and individual spontaneous evaluation.
      (d) Clarifying the dual function of citation and reprint-based evaluation behaviors, which include both qualitative and quantitative factors, and distinguishing these from references.

   **(2) Perspective Innovations**
      (a) Inferring the Evaluation Purpose Constraint Theory, which identifies evaluation purpose as the most critical factor among the six elements. A lack of clear evaluation purpose or misalignment with other elements is the primary cause of the failure of evaluation. The evaluation purpose dictates the selection of subject, object, method, standard, and mechanism. Under specific evaluation purpose, cross-category comparisons—such as between Shakespeare and Picasso—become feasible. Furthermore, it enables the coexistence of common and individualized indicators, as well as the integration of quantitative and qualitative metrics.
      (b) Proposing a three-dimensional evaluation framework wherein any academic entity can be assessed through formal, content, and utility evaluations. This framework progresses from superficial to in-depth analysis, increasing in both complexity and explanatory power compared to traditional binary (quantitative vs. qualitative) evaluations.
      (c) Asserting that rigorous, scientific, and authoritative academic evaluation must be grounded in factual judgment. However, factual judgment alone is insufficient for value judgment.

(d) Recognizing that citation-based metrics (e.g., impact factor, h-index) possess both qualitative (cited by peers) and quantitative dimensions (Ye, 2005). These should be utilized appropriately rather than treated as absolute indicators (Ye, 2010a). While bibliometric methods cannot replace peer review, they provide useful references, validations, and benchmarks, particularly for macro-level comparisons or assessing research productivity.

(e) Distinguishing between academic quality evaluation and academic impact evaluation—with the former mainly focusing on content and utility assessment, using events and cases as primary data sources.

(f) Advocating for the Peer Expert Dominant Theory in quality and specialized evaluations, emphasizing the importance of meta-evaluation (credibility evaluation of reviewers) and appeal mechanisms to address the inherent limitations and subjectivity of peer review.

(g) Arguing that a well-structured evaluation system promotes academic integrity and research ethics.

(h) Positioning academic evaluation as a critical component of Library and Information Science (LIS), encompassing core journal selection, citation analysis, and web-based resource evaluation. As an interdisciplinary field, academic evaluation represents a new growth area for LIS research.

**(3) Theoretical Innovation**

SAER pioneers an all-round analytical framework that integrates all evaluation elements, perspectives, and processes, making it applicable to diverse academic evaluations. SAER extends the traditional binary (qualitative vs. quantitative) evaluation model to a three-dimensional system of formal, content, and utility evaluation framework. This framework not only synthesizes commonalities across output, researcher, and institutional evaluations but also allows for discipline-specific adaptations. It underscores the distinctive features of humanities and social sciences while maintaining connections with evaluation frameworks in natural sciences.

**(4) Methodological Innovation**

SAER integrates bibliometric analysis with expert surveys, case studies, inductive and deductive reasoning, and both analytical and abstract methodologies. Treating all evaluation objects as "information entities"," it applies information methods to construct an academic evaluation system grounded in syntax, semantics, and pragmatics—thus forming a "Complete Information Evaluation Method".

## 4. Implementation and Application of SAER

*4.1 Core Requirements for Implementing*

SAER is applicable across various domains, including the evaluation of academic talent, academic outputs, academic institutions, academic media, and disciplines. The usage requirements are as follows:

**(1) The Evaluation Purpose Should Be Singular**

SAER places significant emphasis on the primacy of the evaluation purpose. It advocates that each evaluation should focus on a single, well-defined purpose based on practical conditions to ensure the accuracy and scientific validity of the conclusions. Empirical evidence has shown that when the evaluation purpose differs, the same evaluator may reach varying conclusions about the same object. Conversely, the same evaluation result may hold different meanings for different subjects being evaluated.

**(2) The Evaluation Process Should Be Diachronic**

SAER recognizes that both the evaluation subject and object are constrained by time and space. The evaluation of the object's utility by the evaluator is not only based on historical facts but must also account for long-term practices and future development trends. Thus, evaluation is a diachronic process that combines both the present (ongoing) and the future (incomplete), requiring continuous validation through practical testing.

**(3) Evaluation Standards Should Be Targeted**

In the SAER, evaluation standards are the value criteria applied to the object during the evaluation process. These standards reflect the evaluator's value orientation and serve to guide the subject being evaluated. For different objects, the evaluation system should be correspondingly optimized and adjusted to suit their specific characteristics.

**(4) Evaluation Methods Should Be Combined**

SAER transcends the binary opposition of evaluation methods. It asserts that evaluation subjects and objects should be selected based on the specific evaluation purpose. Evaluation dimensions should be flexibly combined according to the research difficulty to avoid overly formalized, quantitative, or one-sided evaluation processes.

*4.2 Exemplary Practices of SAER in China*

Over the past decade, SAER has gained significant attention in China's academic and management sectors. It was featured as a topic of discussion at a national evaluation conference (School of Information Resource Management RUC, 2021) and has seen widespread application. Researchers have applied SAER in areas such as scholar, academic journal, specialized database, professional website, academic library collection, research services, librarian, and creative competitions evaluations (Zhao et al., 2020), with positive feedback in practice.

**(1) Book Evaluation**

In 2013, Professor Ye's team applied SAER to evaluate humanities and social science academic books, establishing an All-round evaluation model for Chinese academic books (Wang & Ye, 2014). Through bibliometric methods and expert reviews, they identified high-quality books as representative works and used them to develop the Chinese Book Citation Index (CBkCI) demonstration database (Ye, 2014). This database improved library collection quality, promoted standardized book publishing, and helped researchers quickly identify quality works, laying a foundation for fair and transparent academic evaluations (Ye, 2013b). Ye later refined this model with a more comprehensive framework for evaluating academic book quality (Ye, 2020a).

**(2) Journal Evaluation**

Before the formal introduction of SAER, Professor Ye's team have used its principles to identify 2,770 academic social science journals from 4,500 Chinese journals (Ye et al., 2008). They later evaluated the performance of the Ministry of Education's "Outstanding Academic Journals Program" and laid the groundwork for funding 200 academic journals representing national-level standards (Ye & Zhu, 2024). In 2012, at the request of the Library Society of China, Ye applied SAER as an independent third-party evaluator for selecting outstanding journals in library science (Ye, 2013a). The journals selected received widespread recognition in both academic and industry sectors. SAER also contributed to the establishment of a quality evaluation system for Chinese academic journals (Ye & Yuan, 2021), and it has been used to evaluate non-core Chinese philosophy and social science journals (Zang & Ye, 2018) and the publishing quality of humanities and social science journals in universities (M. Han, 2021).

**(3) Library Evaluation**

SAER has been widely applied in library evaluations. Most notably, Professor Ye's team has created a logical and adaptable quality evaluation framework for academic libraries in China, referencing ISO standards and global evaluation systems for academic libraries (Ye et al., 2021). The team will implement a comprehensive quality assessment of Chinese academic libraries this year. Additionally, based on SAER, Zhejiang University Library, assessed their collection of foreign-language humanities and social science books (Z. Han et al., 2021), and Tianjin Foreign Studies University Library evaluated its subject librarians (Fan, 2012). Besides, Nanjing Engineering University Library developed a framework for evaluating university library print collections (Li, 2020), and the National Library of China created an evaluation system for library exhibition services (Ma, 2020). Nanjing Library also used this for library research and learning activities (Qin, 2024).

**(4) Website Evaluation**
Professor Ye's team used SAER to evaluate Chinese humanities and social science websites, helping researchers access high-quality academic information. Additionally, Lu et al. (2018) evaluated Chinese library websites, and Zhang et al. (2013) assessed websites dedicated to Chinese calligraphy and painting, all contributing to the improvement of website quality and the provision of better information services.

The representative cases demonstrate that the SAER exhibits robust applicability across diverse domains and has garnered substantial practical recognition. Guided by SAER, it becomes feasible to construct research on multiple evaluation objects that demonstrates strong logical coherence, explanatory power, and predictive capability.

## 5. Conclusion

SAER is a systematic framework developed through a comprehensive examination of global evaluation practices in recent years, and has been recognized as one of China's top ten original theories in philosophy and social sciences(Information Center for Social Science Renmin University of China, 2025). It is characterized by the following distinctive features: **(a)** Drawing on research findings from natural science evaluation while emphasizing the unique attributes of humanities and social sciences and their evaluation principles (X. Yang, 2023); **(b)** balancing the universality of academic evaluation with the diversity and specificity across disciplines, fields, and research outputs (S. Yang & Jiang, 2023), reconciling the characteristics of different evaluation domains; **(c)** providing a relatively stable analytical framework while allowing for dynamic development (Jiang, 2023); and **(d)** offering a structured analysis of the historical trajectory of academic evaluation, a clear explanation of its current status and challenges, and a general forecast of its future directions (Geng & Zhao, 2023), bridging the gap between history and the future.

Over the past fifteen years, the SAER has pioneered a comprehensive path in addressing the global challenge of bridging the gap between evaluation theory and practice. In fact, when the 2012 *DORA* and the *Leiden Manifesto* published after STI 2014 called for "indicators must not substitute for informed judgement", SAER had already proactively established a systematic evaluation theory that interlinks aspects such as evaluation purpose. This system, built on six elements and three dimensions, resonates theoretically with principles like the *Leiden Manifesto*'s "Measure performance against the research missions…" (Hicks et al., 2015). It also showcases systematic and dialectical Chinese experience through theoretical inferences like the *Peer Expert Dominant and Meta-Evaluation Theory* and the *Evaluation Mechanism Guarantee Theory*.

Looking to the future, just as scientific research strives to uncover the truth, the construction and refinement of academic evaluation must evolve with the times. Key challenges and opportunities will emerge in addressing the relationship between Chinese characteristics and internationalization, and how to cope with new technologies like generative AI. We wish that SAER will continuously improve through its widespread application in the international academic community, driving the progress of global academic evaluation systems, promoting the high-quality output of scientific research, and ultimately contributing more to academic prosperity, knowledge accumulation, and social progress.

**Open science practices**
In this study, the case data primarily originate from publicly accessible academic databases and published scholarly works, including CNKI (China National Knowledge Infrastructure), CSSCI (Chinese Social Sciences Citation Index), other institutional repositories like Renmin University, and academic monographs authored by Professor Ye (see references for citations). These datasets are published and can be accessed upon purchase. Regarding software, the free version of Xmind was utilized to visually construct the theoretical framework. The use of this software falls within its commercial licensing terms. Additionally, no formal preregistration of the research plan was conducted; however, the theoretical framework and methodology were iteratively designed and thoroughly documented throughout the research process.

**Acknowledgments**
The authors would like to express their gratitude to Dr. Robin Haunschild for his interest in this subject matter and for his gracious invitation, which has significantly contributed to this research.

**Author contributions**
Yu Zhu: Conceptualization, Investigation, Writing – original draft, Writing – review & editing
Jiyuan Ye: Supervision, Funding acquisition, Validation, Writing – review & editing

**Competing interests**
The authors certify that there's no affiliations with or involvement in any organization or entity with any financial interest or non-financial interest in the subject matter or materials discussed in this manuscript.

**Funding information**
This work was funded by the major project of National Social Science Foundation of China (Grant No. 24&ZD323).